\documentclass[reprint,
 amsmath,amssymb,
 aps,
]{revtex4-2}

\usepackage{graphicx}% Include figure files
\usepackage{subcaption}
\usepackage[justification=Justified]{caption}
\usepackage{dcolumn}% Align table columns on decimal point
\usepackage{bm}% bold math
\usepackage{xcolor}
\usepackage{tabularx}
\usepackage[utf8]{inputenc}
\usepackage{array}
\usepackage{float}
\usepackage{lipsum}
\usepackage{hyperref}
\usepackage{appendix}
\hypersetup{
    pdfborder={0 0 0}, % Remove the box around links
    colorlinks=true,   % Colored text instead of boxes
    linkcolor=black,   % Color for internal links ons, figures)
    urlcolor=blue,     % Color for URLs
    citecolor=blue    % Color for citations
}
\usepackage{stmaryrd}
\usepackage[none]{hyphenat}
\begin{document}

\preprint{APS/123-QED}

\title{
Unsupervised learning for the systematic identification of nondispersive wave packets in driven helium
% Unsupervised learning on the identification of non-dispersive wavepackets in the driven helium atom
}

\author{Juan M. Scarpetta$^{1,2}$}
\author{Gustavo A. Parra$^{1, 2}$}
\author{Alejandro González-Melan,$^{2}$}
\author{Javier Madroñero$^{1, 2}$}

\affiliation{$^1$Centre for Bioinformatics and Photonics---CIBioFi, %Edificio E20 No. 1069, 
Universidad del Valle, Cali 760032, Colombia}
%\affiliation{$^2$Departamento de Física, Universidad del Valle, Cali 760032, Colombia}
\affiliation{$^2$Quantum Technologies, Information and Complexity (QuanTIC), Universidad del Valle, Cali, Colombia.}

\date{\today}% It is always \today, today,

\begin{abstract}
Nondispersive wave packets in driven helium are long-lived quantum states that follow classical resonant orbits without spreading. Their identification typically requires detailed analysis of phase-space structures and extensive exploration of parameter regimes. In this work, we introduce an unsupervised learning approach to automate the identification of physically relevant states in the driven helium atom. Using a Floquet-based description, quantum states are computed and represented as probability distributions in configuration and phase space, which serve as input to a convolutional neural network that constructs a low-dimensional embedding of the data. Clustering in the embedding space reveals distinct classes of quantum states. By combining geometric analysis, physical parameter inspection, and time-evolution studies, we identify clusters corresponding to frozen planet states and nondispersive wave packets. The method successfully recovers known NDWP regimes without prior labeling, demonstrating that the learned representation captures physically meaningful structures in a systematic and automated manner. These results establish unsupervised representation learning as an effective tool for the systematic analysis of complex quantum datasets.
\end{abstract}

\maketitle

\section{Introduction}

The helium atom is the paradigm in physics for the three-body Coulomb problem \cite{Ostrovsky} consisting of a heavy charged nucleus surrounded by a pair of electrons, resulting in a repulsive interaction between them and leading to correlation phenomena  \cite{Piraux2003_ProbingEE, Adiabatic}. This system has been extensively studied over the years using classical  \cite{Schlagek_Poincare, Richter-1990}, semi-classical  \cite{SemiclassicalHelium_Wintgen}, and quantum treatments  \cite{Madronero2008}. These studies have provided spectral properties  \cite{madroñeroPhdThesis, eiglesperger2009}, regularization techniques for solving its classical dynamics  \cite{Ostrovsky_AdiabaticTheory}, numerical methods for exact diagonalization  \cite{AG19} and theoretical analyses of the multiple configurations that the atom exhibits  \cite{Schlagek_Poincare}. In general, the dynamics of the helium atom are not regular and display chaotic behavior  \cite{SemiclassicalHelium_Wintgen, Schlagek_Poincare}, except for some stable regions in phase space. This has led to particular interest in studying its stable configurations under constrained dimensionalities, such as the two-dimensional planar helium  \cite{Gonzalez2020, Madronero2008} or the one-dimensional collinear  \cite{Schlagheck_Buchleitner_Collinear} configuration.

In the specific case of the asymmetric collinear $Zee$ configuration, it has been shown that its frozen-planet states (FPS) are stable with respect to ionization energies  \cite{mihelic07} and exhibit long lifetimes  \cite{Schlagek_Poincare, Schlagheck_Buchleitner_Collinear}. These states are characterized by the rapid oscillations of the inner electron, while the outer electron remains nearly frozen around a certain equilibrium position. For all these reasons, FPS are of particular interest as they can be controlled using external fields  \cite{JhonThesis}. When the FPS is periodically driven, it exhibits new properties that can be analyzed through its associated Floquet states. Under near-resonant periodic driving, these states transform into non-dispersive wave packets (NDWPs), which exhibit strong localization along classical trajectories and can be stabilized by electrostatic fields for hundreds of cycles of the periodic driving  \cite{AlejandroThesis}. Recent work~  \cite{Gonzalez2025} has successfully characterized NDWPs in three-dimensional helium by analyzing the sensitivity of the system to driving frequency, field amplitudes, and stabilization via electrostatic fields within the FPS framework.

In general, the dynamics of helium configurations can be studied using spectral expansion methods for configuration-type problems  \cite{SpectralMethod_Generalities} and numerically exact diagonalization techniques  \cite{AlejandroThesis, Gonzalez2025} using Fourier expansions and complex rotation methods  \cite{ComplexRotationMethod}. These computational approaches allow for the calculation of Floquet states and can be implemented numerically~  \cite{AlejandroThesis, Gonzalez2025} across a wide range of parameters—such as total angular momentum $L$, driving frequency $\omega$, field amplitudes and complex energies— yielding an extensive set of quantum states.

Precisely due to this vast parameter space, identifying non-dispersive wave packets is an arduous task that requires extensive analysis of system energies, spectral parameters, field parameters and the lifetimes of excited states. Although all these quantities can be numerically characterized  \cite{JMadroñero_DecayRates, AngulMoment_Contributions}, NDWPs remain hidden within this multidimensional space, and no single parameter combination provides a definitive and sufficient criterion for their identification. Instead, they can be unambiguously identified through their localization properties in both configuration and phase spaces.

Due to the high dimensionality of the system, it is not feasible to visualize the complete wavefunctions directly. Therefore, projections of the states onto three-dimensional configuration and phase spaces are necessary. This is achieved via Husimi distributions, enabling the generation of images representing the probability densities of the states and allowing for the observation of their temporal evolution throughout the period of the external field. Following this approach, in the present work we leveraged established numerical algorithms~  \cite{JhonThesis, AlejandroThesis, eiglesperger2009, Gonzalez2025} to solve and visualize 3D helium frozen planet states as probability distributions in both configuration and phase spaces. This enabled the generation of an extensive image dataset containing both dynamic information of the system and the parameters associated with its behavior in both spaces.

Given that NDWPs exhibit specific localization properties, we propose the use of advanced image analysis techniques~  \cite{scan}, computer vision methods~  \cite{CV_ImageClass}, and convolutional neural networks~  \cite{IIC, Goodfellow2016, MechanismFeatureFearningCNN} to reformulate the identification task as one of geometric feature extraction and clustering in an embedded space, following a Representation/Feature Learning framework. This approach enables the classification of different types of states in the dataset and the identification of non-dispersive wave packets solely based on their geometric localization features, without the need to perform a demanding theoretical analysis of the behavior of the system and its dependence of the relevant parameters.\\

The paper is structured as follows: Sec.~\ref{sec2:TheorMethd} presents the theoretical framework of the helium atom, including the spectral representation and driven states. Sec.~\ref{sec:DatasetGen} describes the dataset generation and data augmentation scheme used for training the machine learning model. Sections~\ref{sec:Results} and \ref{sec:Discussion} present the clustering results and their physical interpretation, including the identification of relevant quantum states. Finally, Sec.~\ref{sec:conclusions} summarizes the main findings and outlines future perspectives.

\section{Theoretical Framework}
\label{sec2:TheorMethd}

\subsection{Helium atom under periodic and static fields}
\label{subsec:helium-periodic}

The theoretical and numerical framework used to compute the driven helium Floquet states has been presented in detail in Ref.~  \cite{Gonzalez2025}. Here we summarize only the elements required to define the quantum states and their representations, which form the basis of the machine-learning analysis introduced below.

We consider the helium atom in the infinite nuclear mass approximation, subjected to a linearly polarized monochromatic driving field and a static electric component. In atomic units, the full Hamiltonian reads
\begin{equation}
H(t) = H_0 + (F \cos \omega t + F_{\textrm{st}})(x_1 + x_2)\,,
\end{equation}
where $F$ and $\omega$ are the amplitude and frequency of the driving field, and $F_{\textrm{st}}$ denotes the strength of the static field, both aligned along the $x$-axis. The field-free Hamiltonian is given by
\begin{equation}
H_0 = \frac{\vec{p}_1^{\,2}}{2} + \frac{\vec{p}_2^{\,2}}{2} - \frac{2}{r_1} - \frac{2}{r_2} + \frac{1}{r_{12}}\,,
\end{equation}
with $r_i = |\vec{r}_i|$ and $r_{12} = |\vec{r}_1 - \vec{r}_2|$. The interaction with the field is treated in the length gauge and within the dipole approximation.

To solve the time-dependent Schrödinger equation associated with this system, we employ the Floquet formalism~  \cite{Floquet,Shirley1965}, which transforms the problem into a time-independent eigenvalue equation for the quasienergies and Floquet states. 
A configuration interaction approach based on Coulomb-Sturmian functions~\cite{SpectralMethod_Generalities,eiglesperger2009} is used to diagonalize the field-free Hamiltonian $H_0$, yielding a discrete atomic basis adapted to the correlated two-electron dynamics. 
The complete Floquet Hamiltonian, including the time-periodic and static field couplings, is then represented and solved in this atomic basis. 
This methodology, which combines the construction of a correlated atomic basis with the Floquet representation of the driven problem, has been developed and applied in detail in Ref.~  \cite{Gonzalez2025}, including the numerical treatment of resonant states via complex scaling. Due to the finite basis representation, the continuum part of the spectrum is discretized, giving rise to continuum-like states that must be distinguished from true resonances through their stability under complex scaling.

The spectrum of the resulting Floquet operator contains bound, resonance, and continuum-like states. To isolate the resonances, we use the complex scaling method~  \cite{ComplexRot,ComplexRotationReinhardt,ComplexRotationMethod}, which analytically continues the coordinates into the complex plane, thereby exposing metastable states as square-integrable solutions with complex eigenvalues $E - i\Gamma/2$, where $\Gamma$ corresponds to the decay rate.

To characterize the spatial and dynamical structure of the Floquet states, we construct representations in both configuration and phase space. 
The configuration-space density $|\Psi(x_1, x_2, t)|^2$ provides spatial information of the electronic wave function at a fixed time, while the phase-space structure of the outer electron is visualized using the Husimi distribution~\cite{Buchleitner2elec,Madronero2008},
\begin{equation}
Q(x, p) = \frac{1}{\pi} |\langle x, p | \Psi \rangle|^2\,,
\end{equation}
where $|x, p\rangle$ denotes a coherent state centered at $(x, p)$. 
The Husimi function provides a smooth, positive-definite representation of the quantum state, suitable for comparison with classical phase-space structures such as Poincaré map~  \cite{SemiclassicalHelium_Wintgen, Schlagek_Poincare} of the classical atom. 

In this context, the identification of relevant quantum states, such as nondispersive wave packets, relies on recognizing localization patterns in these representations and their persistence under time evolution. This observation motivates the image-based approach adopted in this work, where configuration- and phase-space distributions are used as input data for an unsupervised learning scheme.

\subsection{Frozen planet states and nondispersive wave packets}

A central element in the present analysis is the classical frozen planet configuration (FPC), a stable and highly correlated arrangement in which both electrons are collinear and located at the same side of the nucleus~  \cite{SemiclassicalHelium_Wintgen,Richter-1990,Schlagek_Poincare}. 
In this asymmetric configuration, the inner electron follows fast, eccentric trajectories, while the outer electron remains \textit{frozen} around an equilibrium position. 
The resulting dynamics is regular and confined, giving rise to a family of long-lived quantum states known as frozen planet states (FPS), which appear as highly asymmetric doubly excited states associated with ionization thresholds starting from $N=3$.

\begin{figure}
\begin{center}
\includegraphics[width=0.4\textwidth]{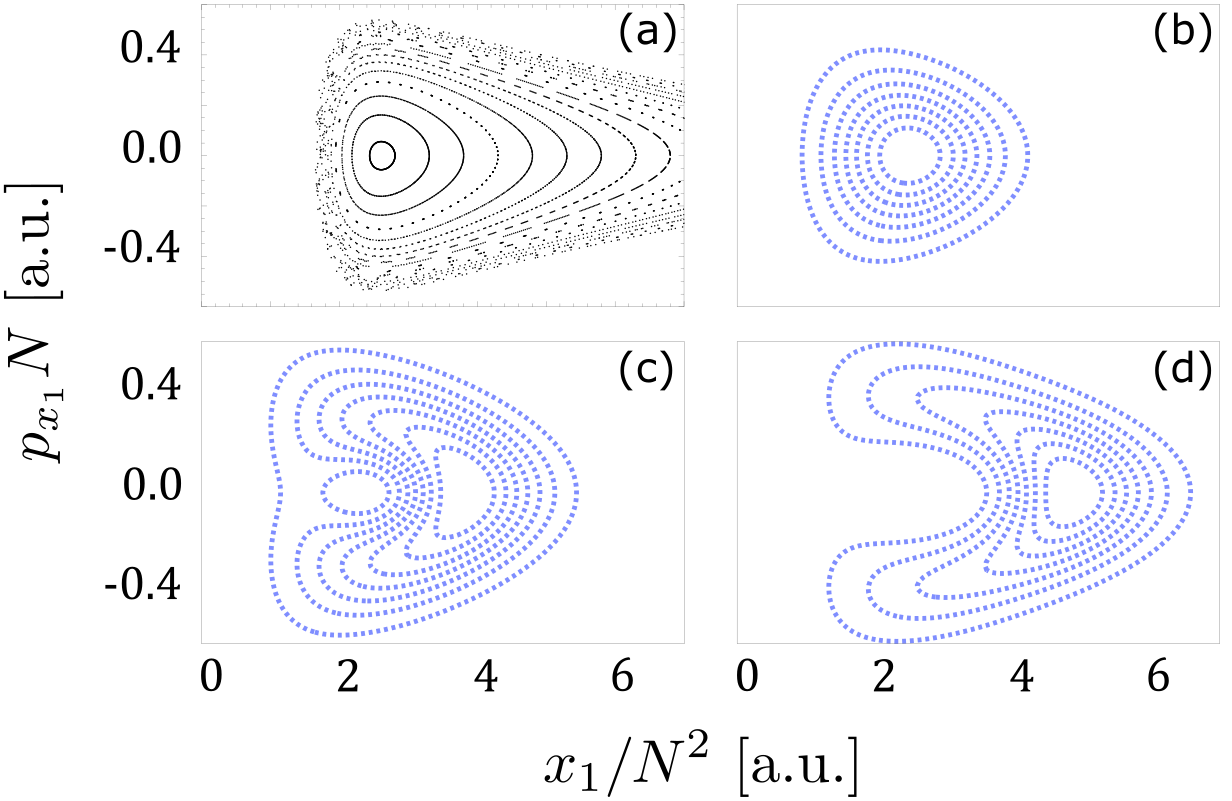}
\end{center}
\caption{Husimi distributions of the ground (b), first excited (c), and second excited (d) frozen planet states in the sixth series. 
The ground state is localized near the equilibrium position of the outer electron. Excited FPS exhibit localization around higher-energy periodic orbits of the classical configuration (a).}
\label{fig:HusimiFPS}
\end{figure}

These FPS can be visualized in phase space through their Husimi distributions. As shown in Fig.~\ref{fig:HusimiFPS}, the ground FPS of a given series is localized near the equilibrium position of the outer electron, 
while excited FPS exhibit phase-space localization around higher-energy periodic orbits of the classical FPC.

In the presence of near-resonant periodic driving, the FPS can give rise to nondispersive wave packets (NDWP)~  \cite{Schlagheck_Buchleitner_Collinear,Madronero2008,Gonzalez2020,Gonzalez2025}, quantum objects that follow the classical periodic orbits of the driven atom without spreading. 
These states can be further stabilized by the application of a weak static field along the polarization axis~  \cite{Schlagheck1998,Gonzalez2025}. 

\begin{figure}
\begin{center}
\includegraphics[width=0.48\textwidth]{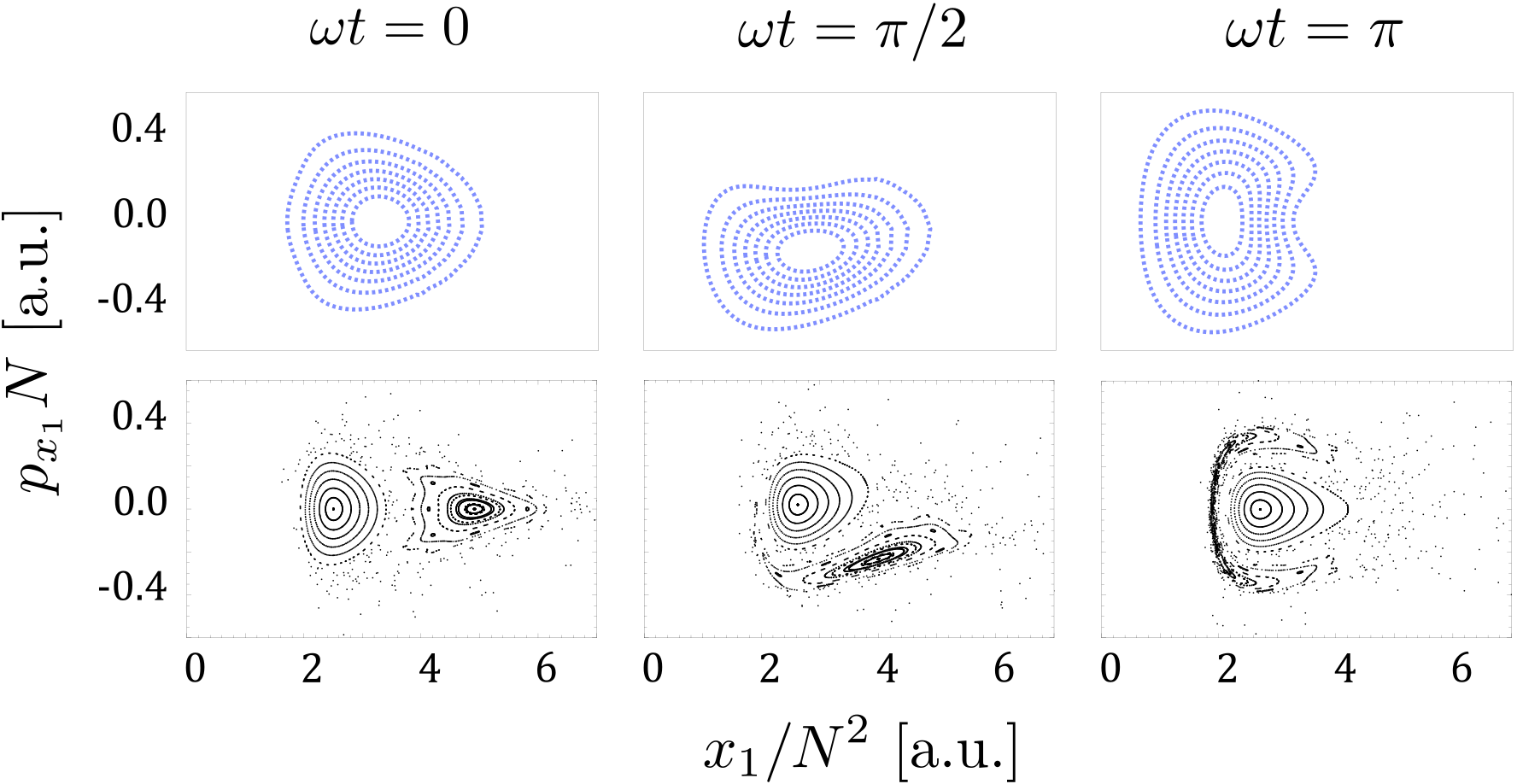}
\end{center}
\caption{Husimi distributions of a nondispersive wave packet below the sixth ionization threshold at different phases of the driving field (top), computed at $F=3.9\times10^{-6}\,\rm a.u.$, $F_{\textrm{st}}=0$, and $\omega=0.00088\,\rm a.u.$. 
The bottom panel shows the corresponding classical phase-space structure for the driven FPC.}
\label{fig:HusimiNDWP}
\end{figure}

In particular, NDWP exhibit phase-space localization within the classical resonance island of the driven dynamics, and remain confined along the corresponding periodic orbit over many cycles of the driving field. 
Fig.~\ref{fig:HusimiNDWP} shows the Husimi distributions of a representative NDWP, illustrating how the wave packet remains localized around the resonance island throughout the driving cycle. 

The identification of NDWP therefore relies on recognizing localization patterns in phase space, as well as their persistence during the time evolution of the driven system. While such identification can be performed through direct inspection of Husimi distributions and spectral properties, this procedure becomes impractical when exploring large parameter spaces. In this work, we address this limitation by reformulating the identification problem in terms of geometric feature extraction from configuration- and phase-space representations, enabling the use of unsupervised learning techniques to classify and identify NDWP automatically.

%%%%%%%%%%%%%%%%%%%%%%%%%%%%%%%%%%%%%%%%%%%%%%%%%%%%%%%%%%%%%%%%%%%%%

\section{Dataset generation}
\label{sec:DatasetGen}

Following the physical framework described above, the identification of nondispersive wave packets is reformulated here as a data-driven problem, where quantum states are represented through their configuration- and phase-space distributions. This requires the construction of a dataset sampling the relevant regions of the parameter space associated with frozen planet dynamics.

The spectral expansion of the wavefunction depends critically on the total angular momentum values $L$ in the superposition. 
In the Floquet representation, the time-periodic problem is mapped onto a discrete set of coupled states, which requires truncation to a finite number of Floquet blocks labeled by $k$ within $k_{\text{min}} \leq k \leq k_{\text{max}}$.
The atomic eigenstates providing dominant contributions to the wave packets are those with angular momentum $0 \leq L \leq 4$  \cite{AlejandroThesis}, where for $L>4$ states contribute near eight orders of magnitude less than $L=0$ states   \cite{AngulMoment_Contributions}. 
Similarly, for the Floquet number $k$ the principal contributions come from $-2 \leq k \leq 2$. 
We quantified these contributions through squared overlaps $|\langle \phi^{LM}_{\alpha} | \Psi \rangle|^2$ between Floquet states and atomic basis states. 
Therefore, our database comprises singlet and triplet states with total angular momentum $L = 0$--$2$ below the 6th ionization threshold and Floquet numbers $-2 \leq k \leq 2$.

Classical analysis of frozen planet states in the $Zee$ configuration~  \cite{Schlagheck_Buchleitner_Collinear}, via adiabatic invariant theory   \cite{Ostrovsky_AdiabaticTheory}, identified optimal driving parameters $F = 1.0 \times 10^{-6}$ a.u. and $\omega = 0.00089$ a.u.~  \cite{Buchleitner2elec} due to pronounced resonant islands~  \cite{Schlagheck_Buchleitner_Collinear} at these values, with electrostatic field $F_{\rm st}$ stabilizing these one-dimensional configurations~\cite{JhonThesis}. Guided by these results, the parameter space was restricted to a region where NDWP are expected to emerge, ensuring that the dataset is physically relevant rather than uniformly sampled over a broad but mostly uninformative domain.

Therefore, our parameter space was scanned systematically within frequencies $\omega = 0.00085$--$0.00093$ ($\Delta\omega=0.00001$ steps), driving fields $F = 1.0\times10^{-6}$--$1.4\times10^{-6}$ a.u. and static fields $F_{\rm st} < 2.23\times10^{-5}$ a.u. (to avoid ionization energies  \cite{JhonThesis}) in steps of $0.1\times10^{-6}$. 
This parameter space guarantees quasi-resonant states consistent with classical dynamics and yields approximately 50 distinct states per time $t$ and per parameter combination, each with characteristic energies and decay rates, resulting in a total of $18~330$ images in the dataset. 

The states in the database $\mathcal{D}$ are labeled with a superscript which represents a unique set of parameters $\alpha \equiv (L_{\text{min}}, L_{\text{max}}, k_{\text{min}}, k_{\text{max}},\omega, F, F_{\rm st}, E )$. 
Each element of the dataset therefore corresponds to a Floquet state associated with a specific point in parameter space.
Therefore, each solution to the TDSE is denoted as $|\Psi^\alpha(t)\rangle$. We evaluate these solutions at different times $t$ within the period $T$ of the driving field as $t=\left\{0, \;T/4, \; T/2\right\}$. 

\begin{figure}[h!]
    \centering
    \includegraphics[width=\linewidth]{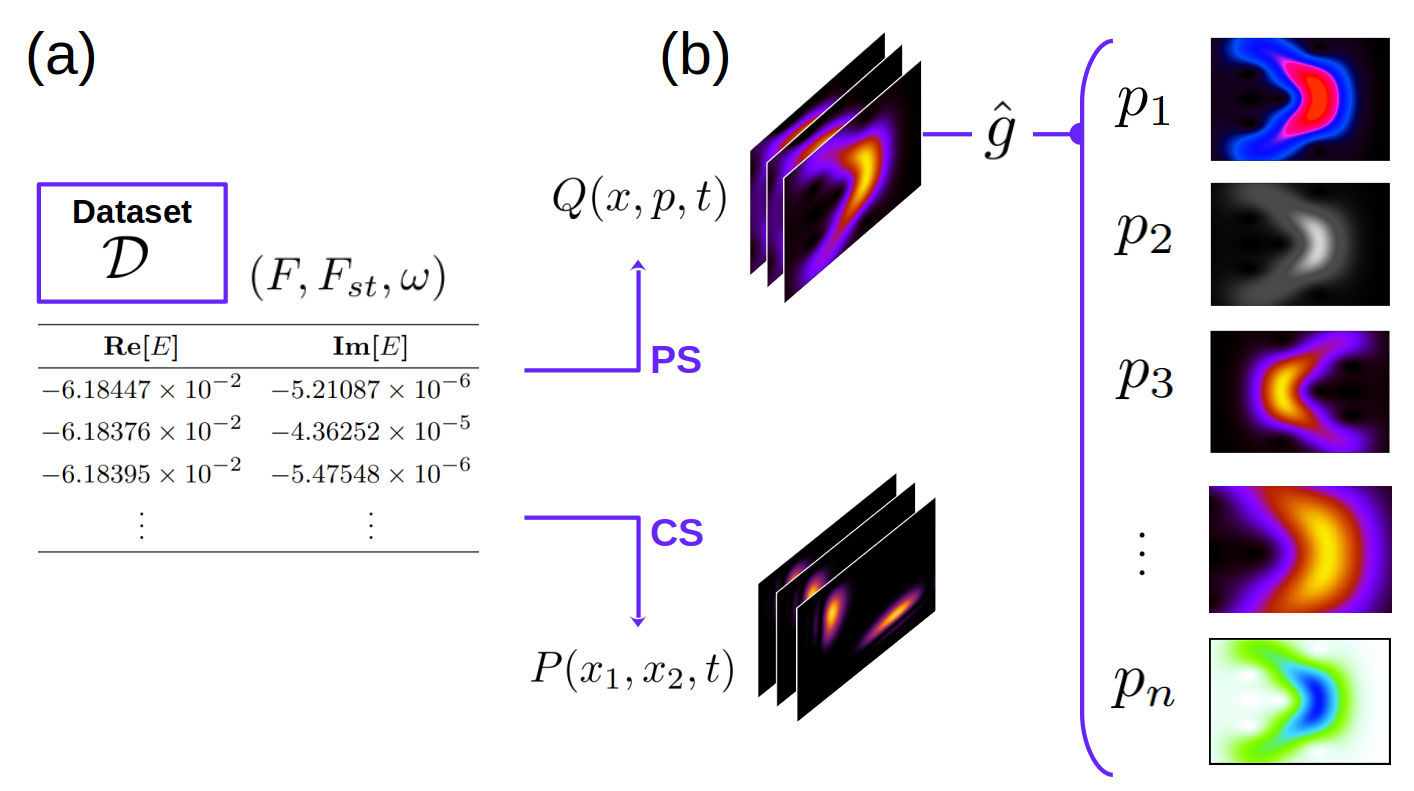}
    \caption{Schematic representation of the generated database. (a) For each parameter combination ($F$, $F_{\rm st}$, $\omega$), approximately 50 distinct states are computed and represented in both configuration space (CS) and phase space (PS). (b) Each image undergoes data augmentation through random transformations $\hat{g}$ sampled uniformly from the complete set of photometric and geometric transformations with probability $p_i$.}
    \label{fig:Dataset}
\end{figure}

Fig.~\ref{fig:Dataset}(a) schematically illustrates the database structure. The essential physical information in our configuration/phase space images is encoded in the probability density patterns along $x_1$, $x_2$, and $p_2$ coordinates, which define the geometric features used to characterize the quantum states.
These representations are invariant  \cite{IIC} under certain image transformations that preserve the geometric properties and can be applied without affecting interpretation of the states. 
These include both photometric transformations (e.g., grayscale conversion, changing contrast and color saturation) and geometric transformations (e.g., scaling, skewing, rotation or flipping). This invariance is a key ingredient for the learning procedure, as it allows the identification of physically equivalent states under different visual representations.

We exploit this through systematic data augmentation~  \cite{dataAugm, dataAugm_Wang} applying random transformations $\hat{g}$ uniformly sampled from these operation families to generate equivalent samples. 
The resulting augmented images serve as equivalent training samples that enhance model robustness   \cite{RobustnessDataAugmentationLoss} and improve the learning process, as illustrated in Fig.~\ref{fig:Dataset}(b).

\section{\label{sec:RL} Representation Learning}

For the training process, the states are represented as images encoded in a rank-3 tensor $x^{(\alpha)}_{\mu \nu \sigma} \in \mathbb{R}^{\text{C} \times \text{W} \times \text{H}}$, where $\text{H}$ and $\text{W}$ denote the height and width of the image and $\text{C}$ is the number of channels. Here, we consider RGB-colored images, thus $\text{C} = 3$. Each image corresponds to a configuration- or phase-space representation of a Floquet state $|\Psi^\alpha(t)\rangle$, as introduced in Sec.~\ref{subsec:helium-periodic}.
In order to perform feature extraction~\cite{MechanismFeatureFearningCNN} of these images, we encode them into a lower-dimensional representation.

\begin{figure}[h!]
    \centering
    \includegraphics[width=\linewidth]{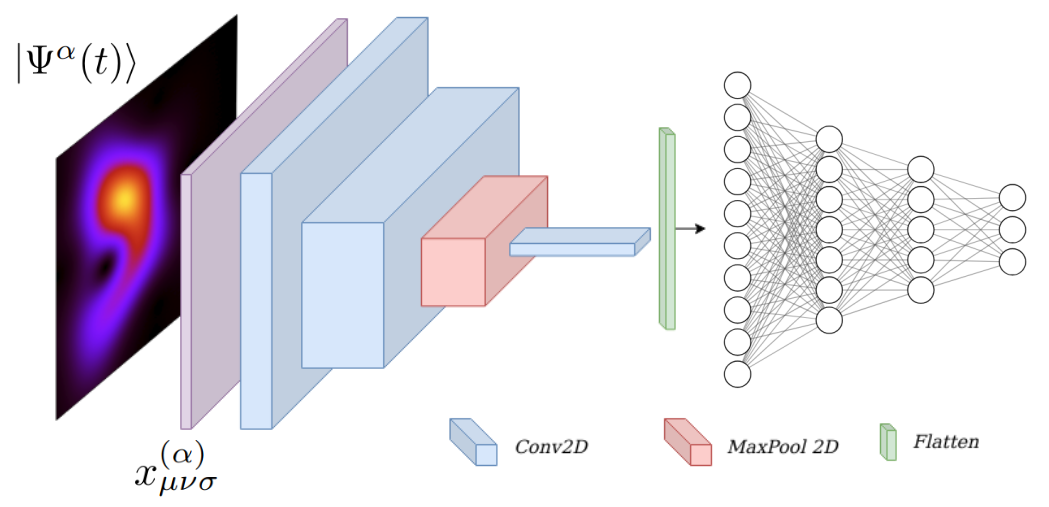}
    \caption{Schematic of the CNN architecture mapping quantum states to $\mathbb{R}^n$ via representation learning.}
    \label{fig:CNN_architecture}
\end{figure}

We construct a convolutional neural network (CNN) $\mathbf{\Phi}_\theta$ which maps the states $|\Psi^\alpha(t)\rangle$ to an $n$-dimensional feature space, i.e.,
\begin{equation}
\mathbf{\Phi}_\theta : x^{(\alpha)}_{\mu\nu\sigma} \mapsto \mathbb{R}^n\,.
\end{equation}
This mapping defines an embedding in which geometrically similar images, corresponding to physically similar quantum states, are expected to be located nearby. The CNN architecture used for this mapping is depicted in Fig.~\ref{fig:CNN_architecture}.

%%%%%%%%%%%% FIGURE OF THE CNN %%%%%%%%%%%%%%%%%
%%%%%%%%%%%%%%%%%%%%%%%%%%%%%%%%%%%%%%%%%%%%%%%%%

The dimensionality $n$ of the output vector constitutes one of the network hyperparameters, along with the kernel sizes, strides, and padding in the convolutional layers, the max-pooling parameters and the activation functions between layers. 
In this way, each image in the dataset $\mathcal{D}$ is represented by a point in the embedding space $\mathbb{R}^n$, allowing us to define a distance function $d(x,y)$ between images, which we take to be the Euclidean metric.\\

The objective of the training process is to extract features that capture the physically relevant structures in the images, such as localization patterns in configuration and phase space, while being invariant under transformations that do not alter the physical interpretation of the state. To achieve this, we minimize the objective function
\begin{equation}
\mathcal{L}=\sum_{\alpha}d\left(\mathbf{\Phi}_\theta\left(x_{\mu\nu\sigma}^{(\alpha)}\right), \mathbf{\Phi}_\theta\left(\hat{g}\hspace{0.8mm}x_{\mu\nu\sigma}^{(\alpha)}\right)\right) +\lambda\sum_{c=0}^n \mathbf{\Phi}_c' \log \mathbf{\Phi}_c'
    \label{Eq:RepresLossFunct}
\end{equation}
where
\begin{equation}
     \mathbf{\Phi}_c' = \frac{1}{|\mathcal{B}_k|}\sum_{\alpha\in \mathcal{B}_k} \sigma\left[ \mathbf{\Phi_\theta}\left(x_{\mu\nu\sigma}^{(\alpha)}\right)\right]_c\,,
\end{equation}
denotes the $c$-th component of the average feature distribution over the batch $\mathcal{B}_k$, obtained after applying the softmax function to the network output.\\

The first term in Eq.~(\ref{Eq:RepresLossFunct}) enforces consistency between representations of images related by transformations $\hat{g}$, ensuring invariance of the embedding under such operations. The second term corresponds to an entropy regularization, which promotes a uniform use of the feature space across the dataset and prevents the collapse of all representations to a single point. The regularization parameter $\lambda$ balances the contribution of both terms and prevents trivial solutions ($\lambda \to 0$).

For the model implementation, we resized all images to a standard dimension of $150 \times 150$ pixels with three color channels. The convolutional layers were configured as follows:
\begin{itemize}
    \item \textbf{Initial convolutions}: Kernel size of 5, stride of 1, and zero padding.
    \item \textbf{Max-pooling}: Kernel size of 2, stride of 2, and zero padding.
    \item \textbf{Final convolution}: Kernel size of 2, stride of 2, and padding of 1.
\end{itemize}

For the fully connected layers, the $\text{ReLU}$ activation function was used in all layers. Additionally, a softmax activation $\sigma$ is applied to the output in order to evaluate the entropy term~\cite{Softmax} in the loss function in Eq.~\eqref{Eq:RepresLossFunct}. These activation functions are defined as:
\begin{equation}
    \sigma(x) = \frac{e^{x}}{\sum_i e^{x_i}}  \qquad  \text{ReLU}(x) = \max\lbrace x, 0\rbrace\,,
    \label{eq:activ_functions}
\end{equation}
where both operations are applied element-wise.

The trainable parameters of the CNN comprise the kernel weights $\Omega_i$, as well as the weights $W_i$ and biases $b_i$ of the fully connected layers. The parameter set $\theta = \lbrace \Omega_i, W_i, b_i \rbrace$ is optimized to minimize Eq.~\eqref{Eq:RepresLossFunct} using the Adam optimizer~\cite{AutomaticDiff_Pytorch}.\\

\begin{figure}[h!]
    \centering
    \includegraphics[width=\linewidth]{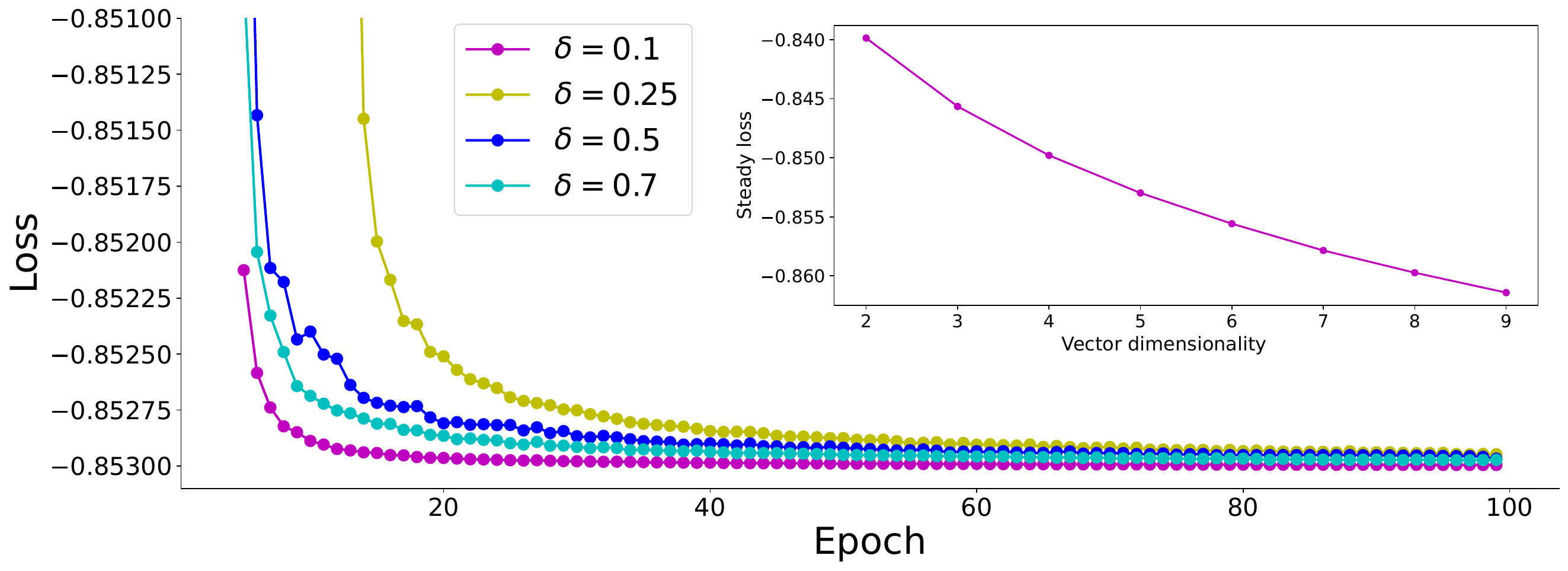}
    \caption{Process of minimizing the cost function as a function of different values of the learning rate $\delta$. Inset shows how the steady loss changes as the $n$ dimension of the output vector increases.}
    \label{fig:Training_Loss}
\end{figure}

Rather than focusing on predictive accuracy, the goal of this training procedure is to construct a meaningful geometric representation of the dataset, in which clusters of points correspond to families of quantum states with similar physical properties. The learning process is shown in Fig.~\ref{fig:Training_Loss}.

The state representation depends on the embedding dimension. As illustrated in the inset of Fig.~\ref{fig:Training_Loss} we evaluated the final loss as a function of this hyperparameter demonstrating that for higher-dimensions $n \geq 10$ the error converges to a steady value, showing no significant improvement compared to lower-dimensional embeddings.

\section{Results}
\label{sec:Results}

%\bf Con los siguientes cambios traté de eliminar algunas redundancias y de introducir una transición más fuerte hacia física dejando claro que clustering es una herramienta y la interpretación es el resultado.

\begin{figure}[h!]
    \centering
    \includegraphics[width=0.82\linewidth]{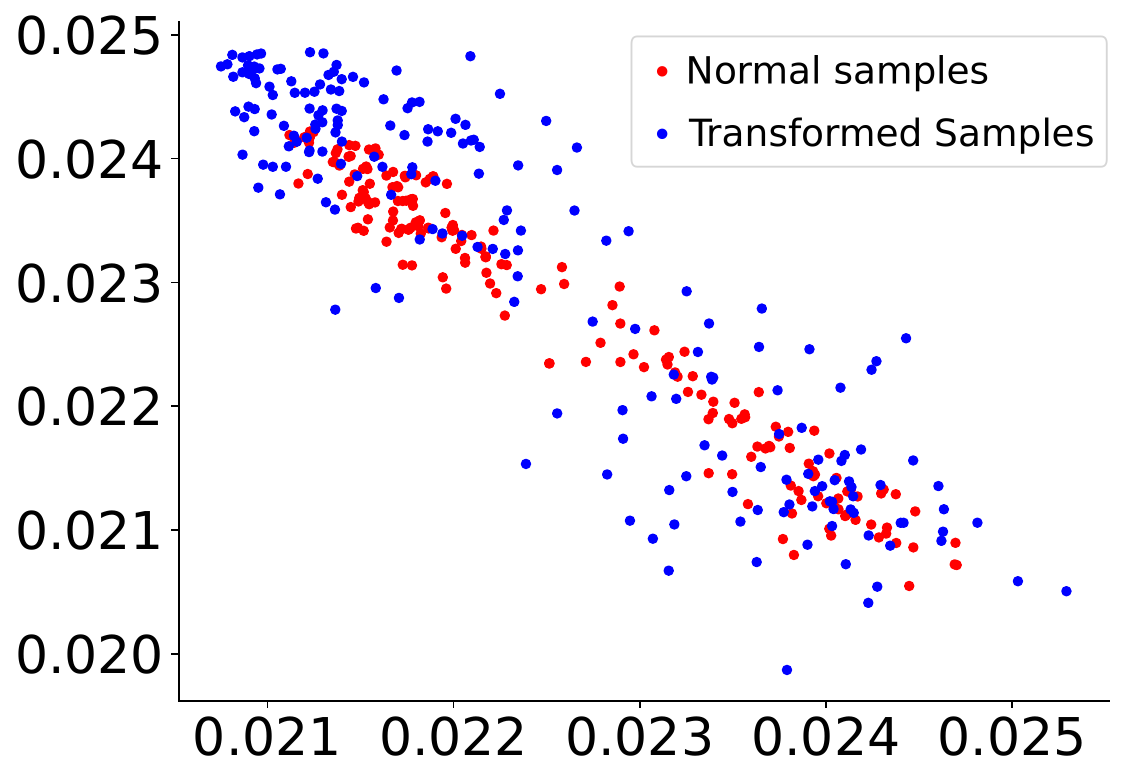}
    \caption{Embedding visualization for a mini-batch of 256 states. The 2D representation shows clear separation between configuration space and phase space states for both original (red) and transformed (blue) samples.}
    \label{fig:n2_clusterization}
\end{figure}

Given the image dataset $\mathcal{D}$, a fundamental first task is to classify states according to their representation in configuration space and phase space. We begin with the minimal embedding dimension $n=2$ for visualization purposes. Fig.~\ref{fig:n2_clusterization} presents the clustering results for this case. The embedded space successfully separates the states into two distinct clusters corresponding to these physically different representations.

\begin{figure}[h!]
    \centering
    \includegraphics[width=0.9\linewidth]{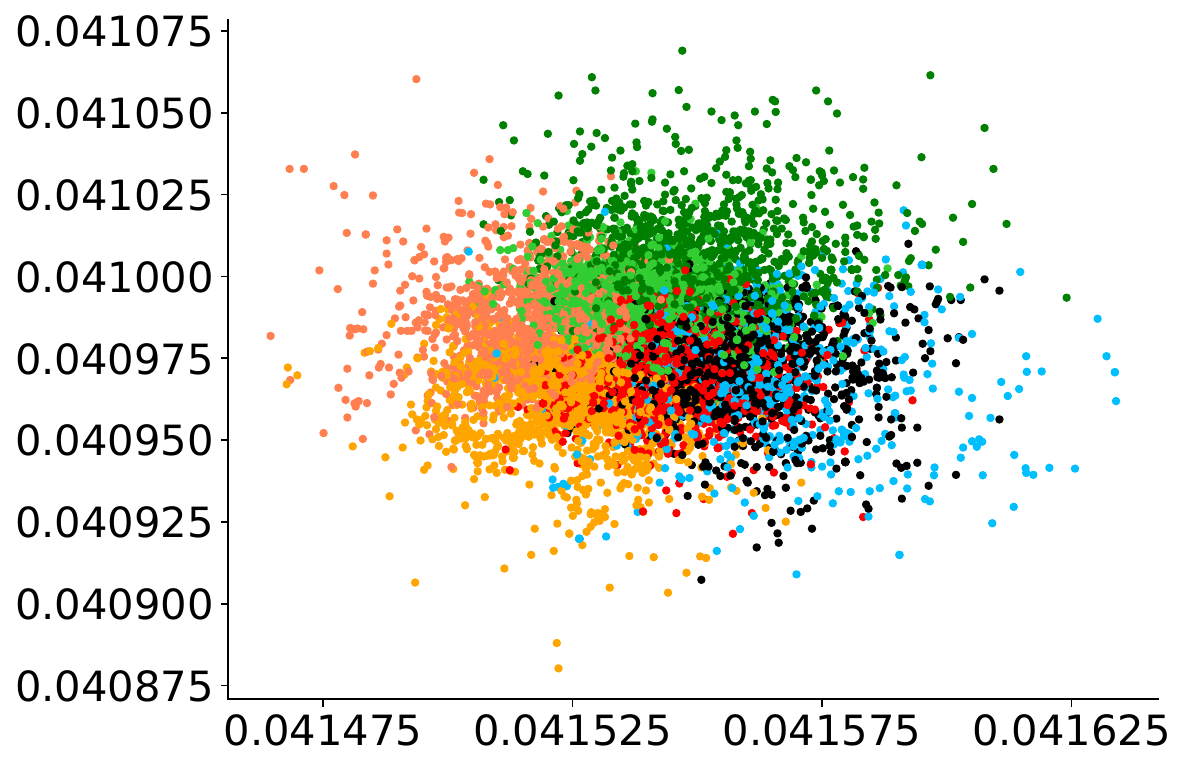}
    \caption{Five-dimensional state clustering (first two components shown) demonstrating separation into $K=7$ distinct classes. Color coding corresponds to different classes predicted by the algorithm.}
    \label{fig:n5_clusterization}
\end{figure}

For a more robust classification, we consider a higher-dimensional embedding space with $n=5$.
This representation enables discrimination among a larger number of distinct state classes. To identify and quantify these classes, we employ the unsupervised $K$-means algorithm~\cite{KMeans_Salman_2011, KMeans_Kanungo}, which determines cluster centroids up to a tolerance of $\epsilon = 10^{-8}$. Fig.~\ref{fig:n5_clusterization} shows the clustering results in the 5D embedding, projected onto the first two principal components. The algorithm reveals $K=7$ well-separated classes at this level of representation, and the cluster structure remains stable in the higher-dimensional space.

Determining the number of clusters is a critical step in the $K$-means procedure, as the hyperparameter $K$ controls the granularity of the classification~\cite{KMeans_Review}. 
We estimate the optimal number of clusters and evaluate cluster quality through the silhouette coefficient~\cite{Silhouette, Silhouette_Graphical}, which quantifies both intra-cluster cohesion and inter-cluster separation~\cite{Stop_Using_Elbow, elbow-silho}. 
The metric yields consistently positive values across all tested configurations, with improved scores for increasing $K$.

% Determining the number of clusters represents a critical step in the $K$-means algorithm, as the hyperparameter $K$ directly controls the granularity  \cite{KMeans_Review} of the classification. We determined the optimal number of clusters  \cite{elbow-silho} and evaluated cluster quality using the silhouette coefficient   \cite{Silhouette, Silhouette_Graphical}, which measures both intra-cluster cohesion and inter-cluster separation  \cite{Stop_Using_Elbow}. The metric yielded consistently positive values across all tested configurations, with improving scores as $K$ increased. For our fine-grained classification  in the configuration space and phase space, we selected $K=466$ clusters ensuring both high silhouette scores and non-empty clusters.\\

For a fine-grained classification of the dataset, we select $K=466$ clusters, ensuring both high silhouette scores and non-empty clusters.
After applying the $K$-means algorithm independently to configuration-space and phase-space representations, we obtain the indices of states belonging to each cluster. These indices allow us to retrieve the corresponding physical parameters $\lbrace \omega, F, F_{\rm st}, \text{Re}[E], \text{Im}[E] \rbrace$, visualize representative states, and compare probability distributions across clusters.

% After applying the $K$-means algorithm independently to each of these two spaces, we determined the indices $i$ of states belonging to each predicted class. These indices allowed us to retrieve the physical parameters $\lbrace \omega, F, F_{\rm st}, \text{Re}[E], \text{Im}[E] \rbrace$ characteristic of each cluster, visualize original images of states and compare probability distributions across classes.
% The most significant clustering results for each space are detailed in subsequent subsections.

The clustering procedure thus provides a structured organization of the dataset, enabling a direct connection between geometric features in the embedding space and the underlying physical properties of the quantum states. In the following section, we focus on the most physically relevant clusters, particularly those exhibiting NDWP behavior, and analyze their structure in detail.

% \subsection{\label{sec:C&P_space}Configuration and phase space clustering}

% Our embedding space analysis resulted in 466 distinct clusters, each corresponding to different types of quantum states. A detailed characterization of all of these clusters is provided in Sec.~\ref{sec:repo}, whereas in Sec.~\ref{sec:Discussion} is shown the representative clustering derived from the configuration space states, emphasizing the most physically meaningful clusters, particularly those exhibiting NDWP behavior for an in-depth analysis. 

\section{Analysis and Discussion}
\label{sec:Discussion}
%\bf JMA: he tratado de reforzar la interpretación física, hacer explícito qué aporta el ML, y evitar que parezca solo “visualización + clustering”

The classification results demonstrated the organization of various Floquet-mode states across different parameter regimes, where the CNN $\mathbf{\Phi}$ groups states primarily through their geometric similarities in visual representations, leading to clusters containing qualitatively similar wave packet distributions. 
Importantly, these geometric similarities correspond to physically meaningful structures in configuration and phase space, allowing the clustering to be interpreted in terms of underlying dynamical behavior.

\begin{figure}[h!]
    \centering
    \includegraphics[width=\linewidth]{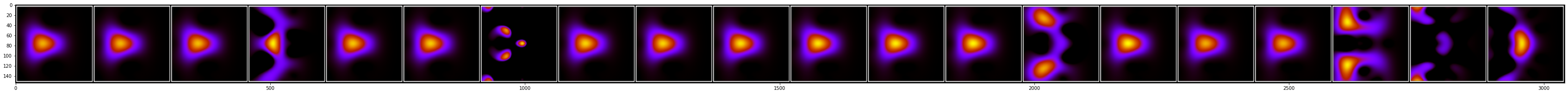}
    \includegraphics[width=\linewidth]{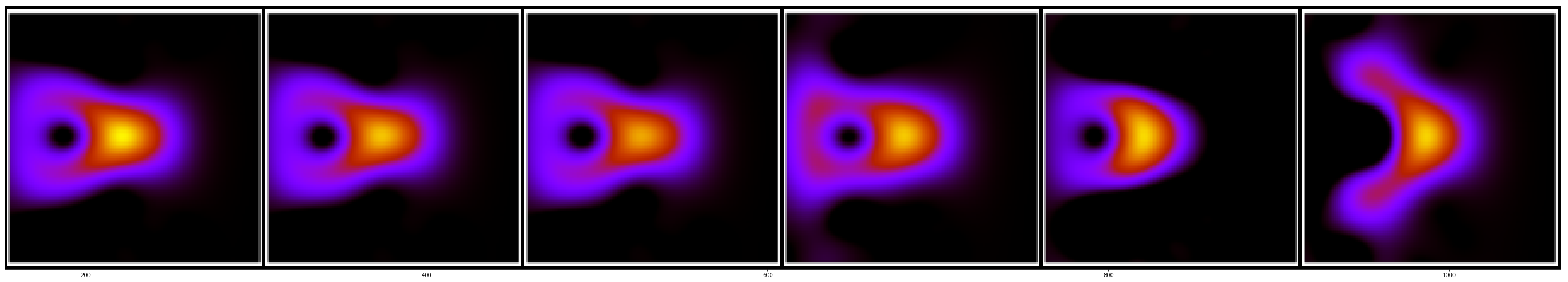}
    \includegraphics[width=\linewidth]{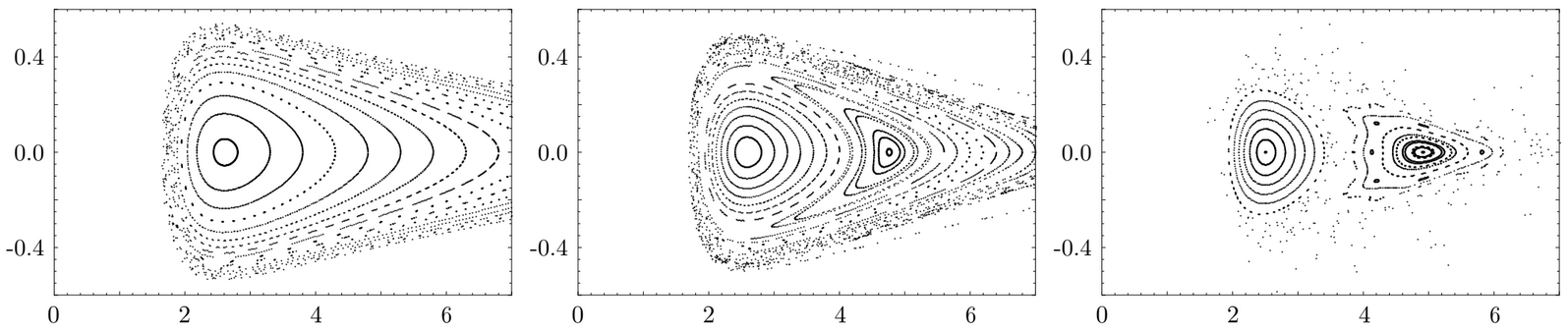}
    \caption{Phase space clustering comparison: (Top) Quantum probability distribution of the cluster with frozen planet states with $\omega=0.00089$; (Bottom) Classical Poincaré sections showing resonance islands emerging at the same frequency $\omega$ for varying field amplitudes $F=0$, $F=0.001N^{-4}$ and $F=0.005N^{-4}$  \cite{AlejandroThesis}}
    \label{fig:fps}
\end{figure}

While the algorithm successfully classifies diverse helium states, our analysis focuses on identifying clusters of well-localized wave packet distributions, particularly those exhibiting the characteristics of frozen planet states~\cite{Schlagheck_Buchleitner_Collinear}. 
In these states, the outer electron undergoes small oscillations around the fixed point, thus maintaining phase-space localization along trajectories associated with resonant classical Poincaré maps~\cite{SemiclassicalHelium_Wintgen}, as shown in Fig.~\ref{fig:fps}. 

Remarkably, the states grouped within this cluster share a common driving frequency $\omega = 0.00089$ and are evaluated at $t=0$, indicating that the clustering procedure is not arbitrary but reflects specific regions of parameter space associated with resonant dynamics. This frequency matches the appearance of resonance islands~\cite{Schlagek_Poincare} in classical Poincaré maps across different field amplitudes $F$, confirming the quantum-classical correspondence in these localized states.

\begin{figure}[h!]
    \centering
    \includegraphics[width=\linewidth]{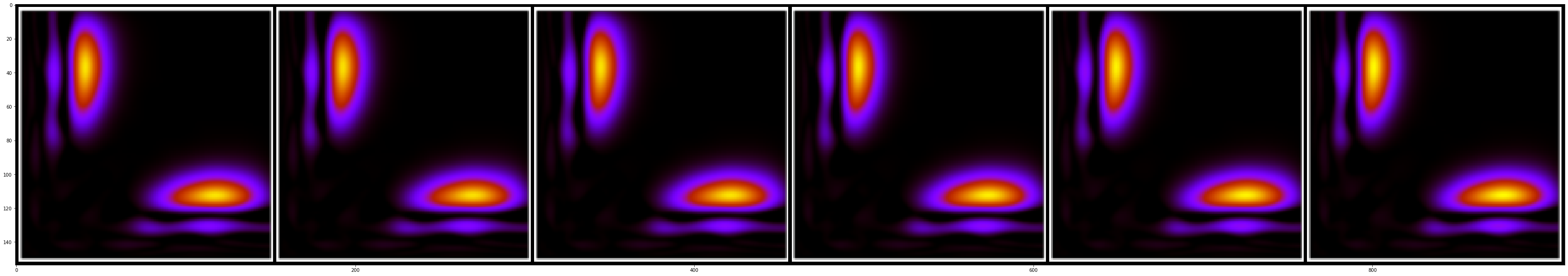}
    \includegraphics[width=\linewidth]{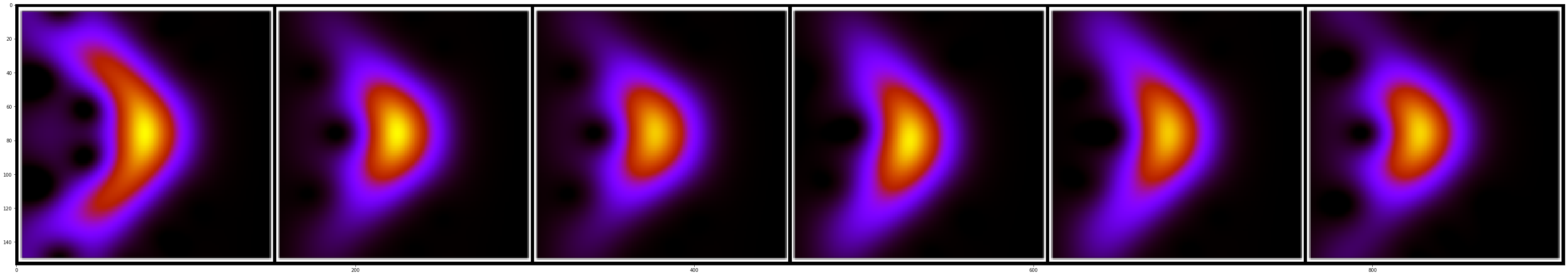}
    \caption{Cluster visualization: The top row shows the probability distributions in configuration space, while the bottom row shows the corresponding representations in phase space.}
    \label{fig:both_results}
\end{figure}

To further validate the physical interpretation of the clusters, we analyze the associated parameters and spatial representations of the states grouped by the algorithm.
Our second identification criterion involved analyzing the physical parameters of states within each cluster while simultaneously visualizing their configuration and phase space representations. 
This dual-space analysis reveals the complete dynamics of the outer electron while simultaneously characterizing the energy scales and decay rates of the states~\cite{liu01, prince05}. 
Fig.~\ref{fig:both_results} displays a representative cluster in both representations, while Table~\ref{tab:cluster_87} quantifies the corresponding parameters.

\begin{table}[h]
\centering
\renewcommand{\arraystretch}{1.2}
\fontsize{9.5}{11}\selectfont
\begin{tabularx}{\linewidth}{>{\centering\arraybackslash}p{0.05\linewidth} % Index
                >{\centering\arraybackslash}p{0.20\linewidth} % F
                >{\centering\arraybackslash}p{0.20\linewidth} % F_st
                >{\centering\arraybackslash}p{0.20\linewidth} % Re[E]
                >{\centering\arraybackslash}p{0.28\linewidth}} % Im[E]
\hline
\multicolumn{1}{c}{\textbf{$\#$}} & \textbf{$F$} & \textbf{$F_{\rm st}$} & Re[$E$]$\times 10^{-2}$ & Im[$E$] \\ \hline
1 & $2.4\times 10^{-6}$ & $1.2\times 10^{-6}$ & $-6.18447$ & $-5.21087\times10^{-6}$ \\ 
2 & $2.5\times 10^{-6}$ & $1.2\times 10^{-6}$ & $-6.18376$ & $-4.36252\times10^{-5}$ \\ 
3 & $2.6\times 10^{-6}$ & $1.4\times 10^{-6}$ & $-6.18395$ & $-5.47548\times10^{-6}$ \\ 
4 & $2.7\times 10^{-6}$ & $1.5\times 10^{-6}$ & $-6.18369$ & $-5.64037\times10^{-6}$ \\ 
5 & $2.8\times 10^{-6}$ & $1.5\times 10^{-6}$ & $-6.18099$ & $-2.00162\times10^{-5}$ \\ 
6 & $2.9\times 10^{-6}$ & $1.4\times 10^{-6}$ & $-6.17618$ & $-8.25326\times10^{-6}$\\ \hline
\end{tabularx}
\caption{Physical parameters for cluster in Fig.~\ref{fig:both_results} states. All states share the resonant frequency $\omega=0.00089$ while varying in fields strength $F$ and $F_{\rm st}$. Energy values show consistent real parts (Re[$E$]) and small imaginary components (Im[$E$]) characteristic of metastable frozen planet configurations.}
\label{tab:cluster_87}
\end{table}

A key step in identifying nondispersive wave packets is verifying the persistence of localization over time. In order to conclusively identify non-dispersive wave packets, we complemented our dual-space analysis by examining the time evolution of states within each cluster. Specifically, we tracked the dynamics at three representative time points $t = \lbrace 0, T/4, T/2 \rbrace$ throughout the period $T$, enabling complete characterization of state evolution and direct comparison with classical predictions.

\begin{figure*}[t]
\centering
\includegraphics[width=\linewidth]{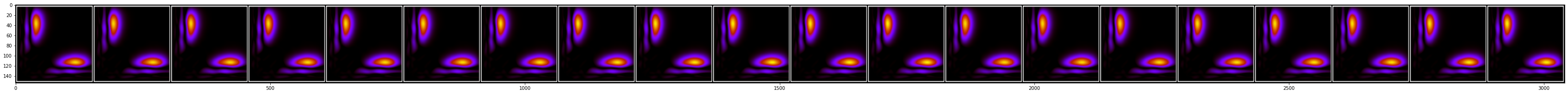}

% \vspace{2mm}

\includegraphics[width=\linewidth]{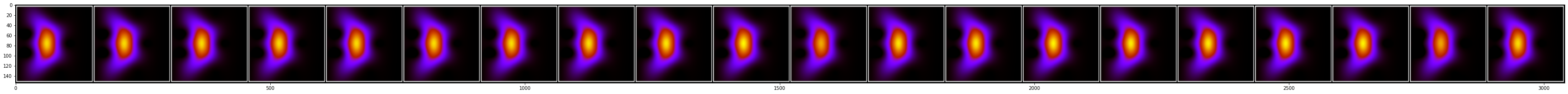}
\includegraphics[width=\linewidth]{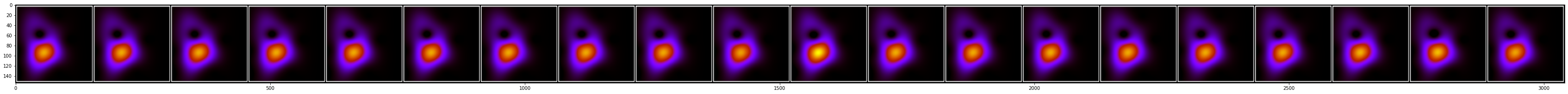}
\includegraphics[width=\linewidth]{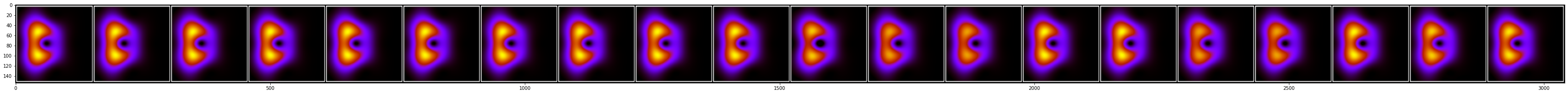}
\captionof{figure}{ Cluster visualization of identified non-dispersive wave packets. (Top) Configuration space representation showing clustered states. (Bottom) Corresponding phase space representations at three evolution times: $t=0$, $t=T/4$, and $t=T/2$.}
\label{fig:final_clusterization}
\vspace{2mm}

%%%%%%%%%%%%%%%%% POSITION 0 - TABLE %%%%%%%%%%%%%%%%%%%%%
\begin{minipage}[h!]{0.49\textwidth}
% \begin{table}[h]
\centering
\renewcommand{\arraystretch}{1.2}
\fontsize{9.5}{11}\selectfont
\begin{tabularx}{\linewidth}{>{\centering\arraybackslash}p{0.05\linewidth} % Index
                >{\centering\arraybackslash}p{0.30\linewidth} % F_st
                >{\centering\arraybackslash}p{0.25\linewidth} % Re[E]
                >{\centering\arraybackslash}p{0.30\linewidth}} % Im[E]
\hline
\multicolumn{1}{c}{\textbf{$\#$}} & \textbf{$F_{\rm st}$} & Re[$E$]$\times 10^{-2}$ & Im[$E$] \\ \hline
1 & $1.0\times 10^{-6}$ & $-6.1720127$ & $-3.25255\times10^{-6}$ \\
2 & $2.0\times 10^{-6}$ & $-6.1720128$ & $-3.25211\times10^{-6}$ \\
3 & $3.0\times 10^{-6}$ & $-6.1720130$ & $-3.25128\times10^{-6}$ \\
4 & $4.0\times 10^{-6}$ & $-6.1720132$ & $-3.25005\times10^{-6}$ \\
5 & $5.0\times 10^{-6}$ & $-6.1720134$ & $-3.24846\times10^{-6}$ \\
6 & $6.0\times 10^{-6}$ & $-6.1720138$ & $-3.24640\times10^{-6}$ \\
7 & $7.0\times 10^{-6}$ & $-6.1720142$ & $-3.24385\times10^{-6}$ \\
8 & $8.0\times 10^{-6}$ & $-6.1720150$ & $-3.24039\times10^{-6}$ \\
9 & $9.0\times 10^{-6}$ & $-6.1720153$ & $-3.23642\times10^{-6}$ \\
10 & $1.0\times 10^{-5}$ & $-6.1720161$ & $-3.23142\times10^{-6}$ \\
11 & $1.1\times 10^{-5}$ & $-6.1720170$ & $-3.22510\times10^{-6}$ \\
12 & $1.2\times 10^{-5}$ & $-6.1720181$ & $-3.21624\times10^{-6}$ \\
13 & $1.3\times 10^{-5}$ & $-6.1720194$ & $-3.20842\times10^{-6}$ \\
14 & $1.4\times 10^{-5}$ & $-6.1720207$ & $-3.19887\times10^{-6}$ \\
15 & $1.5\times 10^{-5}$ & $-6.1720226$ & $-3.18273\times10^{-6}$ \\
16 & $1.6\times 10^{-5}$ & $-6.1720248$ & $-3.16244\times10^{-6}$ \\
17 & $1.7\times 10^{-5}$ & $-6.1720282$ & $-3.13074\times10^{-6}$ \\
18 & $1.8\times 10^{-5}$ & $-6.1720318$ & $-3.09188\times10^{-6}$ \\
19 & $1.9\times 10^{-5}$ & $-6.1720371$ & $-3.03062\times10^{-6}$ \\
20 & $2.0\times 10^{-5}$ & $-6.1720451$ & $-2.92885\times10^{-6}$ \\ \hline
\end{tabularx}
\captionof{table}{Physical parameters showing variation with $F_{\rm st}$ field strength. All states share the resonant frequency $\omega=0.00089$ and fixed field strength $F=1.0\times10^{-6}$.}
\label{tab:NDWP_table}
% \end{table}
\end{minipage}
\hfill
%%%%%%%%%%%%%%%%% POSITION 1 - FIGURE  %%%%%%%%%%%%%%%%%%%%%
\begin{minipage}[t!]{0.49\textwidth}
% \vspace*{0mm}
\centering
\includegraphics[width=1.02\linewidth]{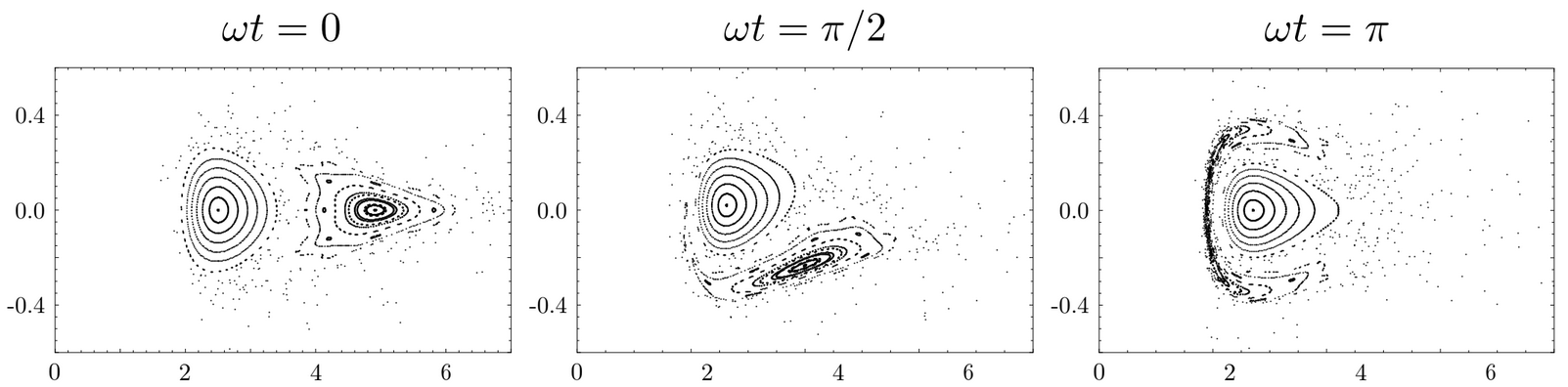}
\caption{Classical phase space of the driven frozen planet configuration for field amplitude $F = 0.005N^{-4}$ a.u., frequency $\omega = 0.2N^{-3}$ a.u. and variable driving field phases $\omega t$. Taken from Ref.  \cite{AlejandroThesis}}
\label{fig:resonant_islands_NDWP}
\vspace{2mm}
\includegraphics[width=1.01\linewidth]{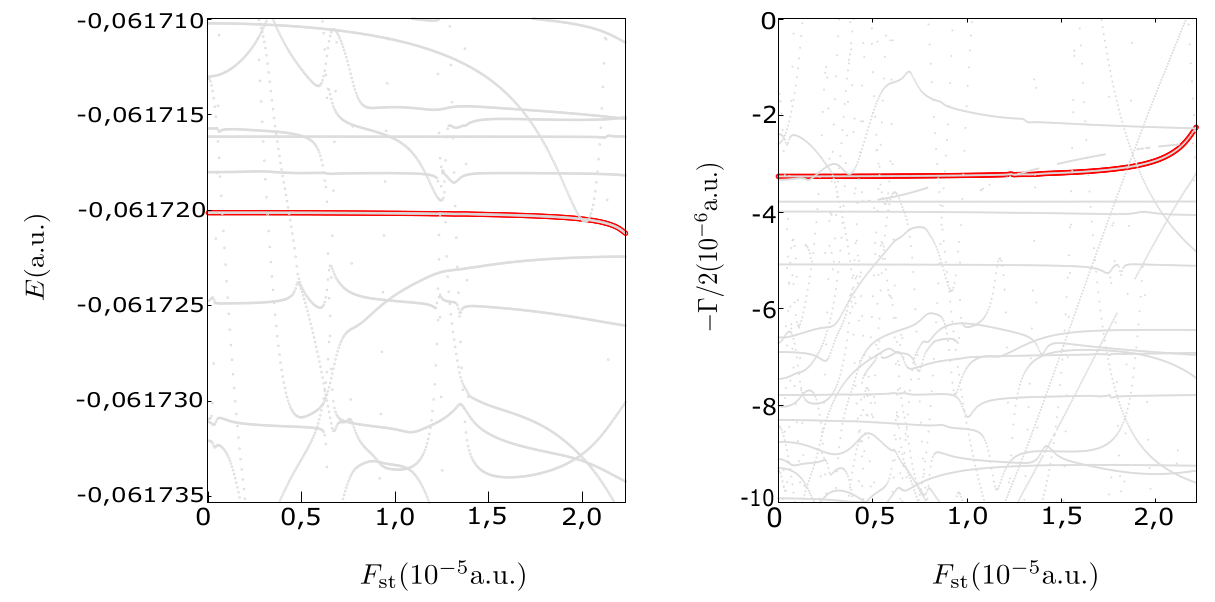}
\caption{Evolution of the real (left) and imaginary part (right) of the energy of the non-dispersive wave packet as a function of the static field at fixed periodic driving with amplitude $F = 1.0 \times 10^{-6}$ a.u. and frequency $\omega = 0.00089$ a.u. Taken from Ref.  \cite{JhonThesis}}
    \label{fig:energies_characterization}
\vfill
\end{minipage}
\end{figure*}

The states belonging to this cluster exhibit a clear signature of nondispersive behavior: their phase-space distributions remain localized along the classical resonance island throughout the driving cycle. Under this criterion, Fig.~\ref{fig:final_clusterization} together with Table~\ref{tab:NDWP_table} shows that this cluster stands out as a representative group for the identification of the states under investigation. This is supported by the geometric consistency observed among the states in configuration space, as well as their phase-space representations, which follow localized distributions along classical trajectories and resemble the dynamics of the classical counterpart, as illustrated in Fig.~\ref{fig:resonant_islands_NDWP}. Furthermore, the parameters associated with these states listed in Table~\ref{tab:NDWP_table} show consistent patterns in terms of resonance energies and decay rates, indicating long-lived states. These values fall within the parameter ranges previously identified for nondispersive wave packets~\cite{JhonThesis, Gonzalez2025}, providing an additional validation of the classification. Altogether, these results demonstrate that the unsupervised learning approach is capable of identifying physically meaningful structures in the dataset, and in particular, of isolating nondispersive wave packets without prior labeling.

\section{Conclusions}
\label{sec:conclusions}

The NDWP states have been previously identified in the helium atom under planar~\cite{Gonzalez2020, JMadroñero_DecayRates} and one-dimensional configurations~\cite{Schlagheck_Buchleitner_Collinear}. 
However, their reliable identification requires a thorough understanding of the classical counterpart, precise characterization of quantum states support on phase space and a comprehensive analysis of stabilization and ionization effects induced by external driven fields, which demands an extensive theoretical investigation. In this context, the identification of NDWPs remains a nontrivial task, particularly when exploring large parameter spaces where manual inspection becomes impractical.

Recent work \cite{Gonzalez2025} has successfully characterized NDWPs in three-dimensional helium by analyzing the sensitivity of the system to driving frequency, field amplitudes and stabilization via electrostatic fields under the FPC framework. Building on these results, the present work introduces a complementary approach based on unsupervised representation learning, aimed at automating the identification of physically relevant states.

Our work leverages established numerical algorithms~\cite{JhonThesis, AlejandroThesis, eiglesperger2009, Gonzalez2025} to solve and visualize 3D helium states as probability distributions in both configuration and phase spaces. We combine these methods with image-based representation learning and clustering techniques, reducing the problem to one of geometric feature extraction in an embedded space. Within this framework, the CNN constructs a low-dimensional representation in which states with similar physical properties are naturally grouped, allowing the application of unsupervised clustering methods. The $K$-means clustering algorithm effectively distinguished multiple classes in both configuration and phase space representations.

While several clusters exhibited potential NDWP characteristics, rigorous validation required inspection of key state parameters, including field amplitudes, driving frequencies, and energy levels. 
This step provides a direct link between the data-driven classification and the established physical criteria for NDWP identification.

Cross-validation with prior studies~\cite{AlejandroThesis,JhonThesis} led to the rejection of some clusters due to inconsistencies in energy scales or system parameters. 
For the most promising candidates, we performed a detailed analysis by visualizing the states at three distinct temporal snapshots ($t = 0$, $T/4$, and $T/2$) in both configuration and phase spaces to confirm their dynamical properties. 

In particular, the cluster shown in Fig.~\ref{fig:final_clusterization} exhibits all defining features of nondispersive wave packets, including phase-space localization along classical resonance islands and persistence over time. The corresponding physical parameters are consistent with previously reported NDWP regimes, providing an independent validation of the method. Although no previously unknown NDWP states were identified within the explored parameter space, the results demonstrate that the proposed approach enables their systematic and automated identification, overcoming the limitations of manual inspection. This represents a key step toward automated classification strategies in strongly correlated and driven quantum systems.

% However, our analysis did not reveal novel NDWP states beyond those previously known, although several states exhibiting partial NDWP characteristics were identified. 

Future research could benefit from a complete characterization of all Floquet states obtained through solutions of the eigenvalue problem, as well as from extending the analysis to higher ionization thresholds. This would enable a direct comparison between clustering-based and traditional parameter-based classifications, potentially facilitating the identification of other physically relevant states beyond NDWPs. 

In addition, incorporating dynamical information directly into the learning framework may further improve the identification of states with nontrivial temporal behavior. Additionaly, alternative criteria and machine learning methods could be implemented to characterize, classify and identify quantum states from different viewpoints, including not only static geometric properties but also dynamical features. Accordingly, various computer vision techniques may be applied, such as trajectory similarity measures~\cite{Trejectory_Measure, Tao_trajectory_measure}, temporal evolution clustering~\cite{Evolution_Clusters}, optical flow analysis~\cite{Masker_OF_2025} and 3D convolutional neural networks~\cite{SpatioTemporal_features3DCNN}, among others.

% Additionaly as a perspective for future work, alternative criteria and Machine Learning methods could be implemented to characterize, classify and identify quantum states from different viewpoints, including not only the static geometric properties of states projected in configuration and phase spaces but also their dynamical features. Accordingly, various computer vision techniques may be applied, such as trajectory similarity measures  \cite{Trejectory_Measure, Tao_trajectory_measure} to quantify the deviation of the distributions from classical orbits, temporal evolution clustering  \cite{Evolution_Clusters}, optical flow analysis~  \cite{Masker_OF_2025} to detect regions of motion in the probability distributions and 3D convolutional neural networks  \cite{SpatioTemporal_features3DCNN} for spatio-temporal feature analysis of the quantum states, among others, with the aim of strengthening the criteria and evaluating the performance of several unsupervised techniques for this specific and important classification task in helium.

%\section{Data and Code Repository\label{sec:repo}}
%\azul{\bf jma: no creo que sea necesaria esta sección. sugiero eliminarla. Faltaría una sección de agradecimientos. No es obligaroria.} The database and Fortran codes used to develop this work are available upon reasonable request to the authors$^{1,2}$.

%% REFERENCES
\nocite{*}
\bibliography{refs}

%%%%%% APPENDIX
% \section{Appendix: ML and clusters\label{app:ML}}

% La clusterización en el espacio embebido dejó \lipsum[1-3]

% \begin{figure}[h!]
%     \centering
%     \includegraphics[width=\linewidth]{Results/Classification_CfSpce/cls_46.png}
%     \includegraphics[width=\linewidth]{Results/Classification_CfSpce/cls_3.png}
%     \includegraphics[width=\linewidth]{Results/Classification_CfSpce/cls_17.png}
%     \includegraphics[width=\linewidth]{Results/Classification_CfSpce/cls_21.png}
%     \includegraphics[width=\linewidth]{Results/Classification_CfSpce/cls_74.png}
%     \includegraphics[width=\linewidth]{Results/Classification_CfSpce/cls_81.png}
%     \caption{Configuration space clustering results showing six representative classes.}
%     \label{fig:config_space_clustering}
% \end{figure}

% \begin{figure}[h!]
%     \centering
%     \includegraphics[width=\linewidth]{Results/Classification_PhSpce/cls_21.png}
%     \includegraphics[width=\linewidth]{Results/Classification_PhSpce/cls_78.png}
%     \includegraphics[width=\linewidth]{Results/Classification_PhSpce/cls_100.png}
%     \includegraphics[width=\linewidth]{Results/Classification_PhSpce/cls_116.png}
%     \includegraphics[width=\linewidth]{Results/Classification_PhSpce/cls_145.png}
%     \includegraphics[width=\linewidth]{Results/Classification_PhSpce/cls_330.png}
%     \includegraphics[width=\linewidth]{Results/Classification_PhSpce/cls_369.png}
%     \caption{Phase space clustering results showing seven distinct dynamical regimes.}
%     \label{fig:phase_space_clustering}
% \end{figure}

\end{document}